\documentclass[fleqn,10pt]{wlscirep}
\usepackage[utf8]{inputenc}
\usepackage[T1]{fontenc}
\usepackage{booktabs}
\usepackage{mathtools}
\usepackage{comment}
\usepackage{multirow}
\usepackage{float}
\usepackage{hyperref}
\usepackage{subcaption}
\usepackage{color}
\usepackage{mathptmx}
\usepackage{dcolumn}

\title{The role of the Big Geographic Sort in the circulation of misinformation among U.S. Reddit users}
\author[1]{Lia Bozarth}
\author[*,2,3]{Daniele Quercia}
\author[4]{Licia Capra}
\author[2]{Sanja Scepanovic}
\affil[2]{Bell Labs, Cambridge, United Kingdom}
\affil[3]{CUSP, Kings College London, United Kingdom}
\affil[1]{University of Michigan, USA}
\affil[4]{University College London, UK}

\affil[*]{quercia@cantab.net}
\begin{abstract}
Past research has attributed the online circulation of misinformation to two main factors -  individual characteristics (e.g., a person's information literacy) and social media effects (e.g., algorithm-mediated information diffusion) - and has overlooked a third one: the critical mass created by the offline self-segregation of Americans into like-minded geographical regions such as states (a phenomenon called `The Big Sort'). We hypothesized that this latter factor matters for the online spreading of misinformation not least because online interactions, despite having the potential of being global,  end up being localized: interaction probability is known to rapidly decay with distance. Upon analysis of more than 8M Reddit comments containing news links spanning four years, from January 2016 to December 2019, we found that Reddit did not work as an `hype machine' for misinformation  (as opposed to what previous work reported for other platforms, circulation was not mainly caused by platform-facilitated network effects) but   worked as a supply-and-demand system: misinformation news items scaled linearly with the number of users in each state (with a scaling exponent $\beta$ ~$\approx 1$, and a goodness of fit $R^2\approx 0.95$). Furthermore, deviations from such a universal pattern were best explained by state-level personality and cultural factors ($R^2\approx \{0.12, 0.39\}$), rather than socioeconomic conditions ($R^2\approx \{0.15, 0.29\}$) or, as one would expect, political characteristics ($R^2\approx \{0.06, 0.21\}$). Higher-than-expected circulation of any type of news (including reputable news) was found in states characterised by residents who tend to be less diligent in terms of their  personality  (low in conscientiousness) and by loose cultures understating the importance of adherence to norms
(low in cultural tightness). Interestingly, the combination of those factors  with low levels of education was then associated with the particular circulation of misinformation. These results suggest that
online interactions are geographically bounded and, as such,  circulation of misinformation cannot be studied purely as an Internet phenomenon but should be grounded into a user's offline cultural environment, which  has become increasingly segregated over the decades, and is admittedly hard to change.

\end{abstract}

\begin{document}

\flushbottom
\maketitle
%
%
\thispagestyle{empty}

\section*{Introduction}
The circulation of misinformation online is a major contributor to various contemporary political and social issues in the U.S., including the Clinton Email and Pizzagate conspiracies in 2016, the immigration crisis (migrant caravans) in 2018, and the January 6th riot in the U.S. Capitol building in 2021. It is also associated with hyper-polarization of opinions, xenophobia, and civil unrest~\cite{gradon2020crime,duong2021covid}. Alarmed by fake news' continued grip on a significant portion of the U.S. population and its persistent negative impacts, researchers have extensively examined factors associated with an individual’s tendency to spread misinformation. Past research has attributed circulation to two main categories of factors. The first category includes individual characteristics such as a person's personality and culture, education attainment, and political-leaning~\cite{forgas2019social,burbach2019shares,buchanan2019spreading,grinberg2019fake,balestrucci2020credulous}. For example, conscientious (careful, diligent) individuals were shown to be significantly less likely to spread false content, whereas credulous individuals tend to be significantly more likely~\cite{forgas2019social,balestrucci2020credulous}. Further, those with lower information literacy were observed to be more likely to spread misinformation~\cite{scherer2021susceptible}. Similarly, past research has determined that individuals also share misinformation in an effort to attack ideological opponents~\cite{osmundsen2021partisan}. 

The second category has to do with the ways social media are engineered to work as a ``Hype Machine''~\cite{aral2020hype}. For instance, existing social media platforms'  ``friends suggestion algorithms''---which tend to disproportionately recommend friends of friends who likely share similar behaviors and beliefs---have amplified the online clustering of individuals into homophilous communities. Prior studies demonstrated that these small and densely connected online communities, in turn, had significantly increased the spread size, depth, and speed of misinformation~\cite{vosoughi2018spread}. For instance, Cinelli showed that users on Facebook are much more ideologically segregated compared to those on Reddit~\cite{cinelli2021echo}. In addition, users were also observed to be more likely to share ideologically congruent low-credibility news content to their immediate network~\cite{guess2021cracking}. This platform-enabled circulation of misinformation is commonly known as the echo chamber or filter bubble effect~\cite{pariser2011filter,jamieson2008echo}. Another platform-amplified feature is affect. Platform algorithms were observed to preferentially recommend emotionally salient and polarizing content to boost user engagement and content sharing~\cite{bakir2018fake,rathje2021out}. Given that misinformation tends to be more sensational and novel, this algorithmic bias had also led to the oversharing of misinformation~\cite{vosoughi2018spread}. 


There is, however, a third overlooked factor: the offline self-segregation of Americans into like-minded communities such as geographic states, a phenomenon which Bill Bishop dubbed as ``The Big Sort''~\cite{bishop2009big}. Work by Bishop and others has illustrated that people in the U.S. have been increasingly choosing to live in neighborhoods populated with others who are just like themselves in values and beliefs. Furthermore, this sorting has resulted in geographical regions (e.g., states) with distinct lifestyle and culture~\cite{glass2014red,monson2011all,jokela2015geographically}, political ideology~\cite{scala2017political}, and even personality~\cite{rentfrow2009statewide,rentfrow2013divided,elleman2018personality}. As an example, work by Rentfrow et al.~\cite{rentfrow2013divided} showed that the states of Utah and New York are the most and least agreeable among all the states, respectively. South Carolina is the most conscientious, and Maine the least. Similarly,  Mississippi has the most restrictive cultural and social norms, whereas California has the most loose~\cite{rentfrow2013divided}. Furthermore, states' personality and culture are indicative of their voting patterns~\cite{rentfrow2009statewide}. The process of Americans geographically sorting themselves over the past four decades into homogeneous communities still continues. Thus far, it is unclear whether it has had any impact on online news circulation, particularly the circulation of misinformation. 

To ascertain that, we examined the geographical circulation of news on Reddit, a popular online content aggregation and discussion website. Reddit consists of many communities (or areas of interest) called subreddits that function akin to online forums. Users can make public posts on these subreddits and others can then comment on the original posts. For instance, a user can post a news article about Covid-19 on the subreddit r/news, and others can then discuss the article with each other. Unlike social media platforms such as Twitter and Facebook, Reddit users are anonymous. We chose Reddit for our analysis given that it has one of the most comprehensive publicly available archived datasets (available under \url{pushshift.io}). Additionally, Reddit was also the home for the notorious misinformation hub: r/the\_Donald, which was a subreddit responsible for spreading a considerable volume of misinformation across other platforms~\cite{zannettou2017web,zannettou2018origins}.

\section*{Data}
\label{sec:data}

\begin{figure}[t!]
    \centering
    \includegraphics[width=0.45\linewidth]{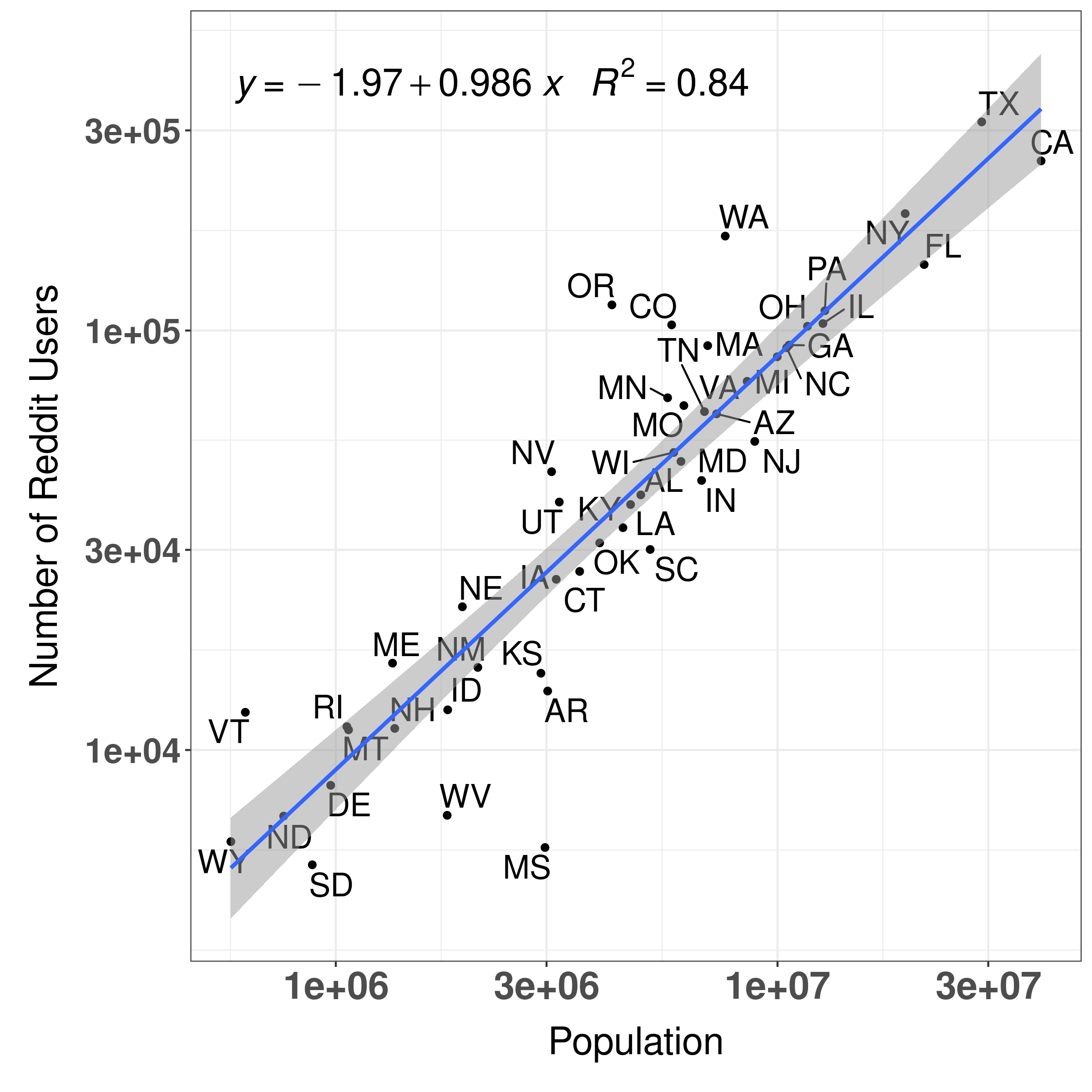}
    \caption{The $x$-axis denotes each state's population (logged) and the $y$-axis is the number of Reddit users from each state. We see that the number of Reddit users scaled linearly with population ($\beta=0.99$), with an $R^2=0.84$.}
    \label{fig:rep_pop}
\end{figure}

\vspace{0.15cm}
\noindent
\textbf{Reddit Data.} We used Pushshift's~\cite{baumgartner2020pushshift} publicly available comments dataset from January 2016 to December 2019. This dataset contained all comments from all public and quarantined subreddits (e.g., r/the\_Donald was quarantined for a time before it was permanently removed on June 29, 2020). We then used the method from Balsamo et al.~\cite{balsamo2019firsthand} to assign users to their geographical location. Specifically, we first identified a list of 2.87K subreddits that can be matched to one of the U.S. states (e.g., r/seattle, r/california). Then, for each user who had posted at least once in these subreddits, we assigned the user to the corresponding U.S. state. Note that if a user had posted in multiple states, we assigned the user the state with the majority of posts. As a result, 82.4\% of users had only posted in a single state, and 95.2\% of users had posted in at most 2 states. Finally, only 3.8\% of users were not assigned a state due to not having a majority state. We identified approximately 3M users who were located in one of the 50 U.S. states. The correlation between a state's population and its number of Reddit users is shown in Figure~\ref{fig:rep_pop}. We saw that the number of Reddit users per state scaled linearly with the state's population ($\beta=0.99$). Additionally, approximately 1.4 billion (or 35\%) comments on Reddit can be mapped to a user in one of the 50 U.S. states. From the 1.4B comments, we identified a total of 8.23M (0.6\%) comments containing news links. Next, we classified a Reddit comment as a {\it fake} news comment if it  contained a URL to a domain that appeared in the list of {\it fake} news domains, as per the groundtruth labeling procedure described next.


\begin{table}[t]
\centering\scriptsize
\caption{Summary Statistics for News Comments. These comments are Reddit posts that contain links to news articles of three types.}
\label{tab:dat}
\begin{tabular}{l|rrrrr}
  \hline
news\_type & unique\_comments & unique\_user & unique\_news\_site & unique\_urls  & top\_news\_sites \\ 
  \hline
fake & 116212 & 45485 & 933 & 60754 & breitbart.com, dailywire.com, thegatewaypundit.com\\ 
  lowcred & 536701 & 160146 & 1801 & 264010 & dailymail.co.uk, washingtonexaminer.com, dailycaller.com\\ 
  reputable & 7645044 & 717198 & 5221 & 3319213 & nytimes.com, washingtonpost.com, wsj.com\\ 
   \hline
\end{tabular}
\end{table}

\vspace{0.15cm}
\noindent
\textbf{Website Groundtruth Labels.} We compiled a list of news websites (or domains) from various sources widely used in researching misinformation~\cite{bozarth2020higher}. Each news site was then labelled as one of three types -  {\it fake, lowcred}, or {\it reputable} - as follows. 

\vspace{0.15cm}
{\it\noindent Reputable:} We used three sources to compile a list of reputable news sites: Vargo et al.~\cite{vargo2018agenda},  Alexa (\url{alexa.com}), and Media Bias/Fact Check (\url{mediabiasfactcheck.com}). This resulted in 8.9k total reputable news sites. 

\vspace{0.15cm}
{\it\noindent Fake:} Based on a detailed meta-review  in related work~\cite{bozarth2020higher}, we compiled a list of questionable news sites from 5 existing sources: Zimdars list~\cite{zimdars2016my}, Media Bias/Fact Check, PolitiFact~\cite{politifact}, the Daily Dot~\cite{dailydot}, and Allcott, et al~\cite{allcott2018trends}. By using the descriptions and granular labels of each of the  five sources, we categorized a domain as {\it fake} if it had routinely published completely fabricated news articles. There were a total of 933 unique fake news sites across all five sources.


\vspace{0.15cm}
{\it\noindent Lowcred:} Unlike {\it fake} news sites, low-credibility news sites publish articles with mixed factualness rather than completely fabricated content. We included domains that were described by the previous 5 sources as unreliable, hyperpartisan, clickbait, rumor, pseudoscience, and conspiracy sites, ending up with a total of 1801 low-credibility news domains.
\begin{table}[t!]
\centering\scriptsize
\caption{List of State-level Attributes. \label{tab:stateattr}}
\begin{tabular}{p{3cm}|p{2cm}|p{8.6cm}}
  \hline
\bf{Category} & \bf{Variable Name} & \bf{Description} \\ 
  \hline
  personality and culture  & openness & imaginative, spontaneous\\ 
  & conscientiousness & disciplined and careful\\ 
  & extraversion & social and fun-loving \\ 
  & agreeableness & trusting and helpful \\ 
  & neuroticism & anxious, pessimistic \\ 
  
  & cultural\_tightness & restrictive social norms and punishments for deviance\\\hline
socio-economic & density  & population density (proxy for urbanization) 2019\\ 
    & gdp & state's gdp per capita 2019\\ 
  & minority & percentage of person of color 2019\\ 
  & no\_highschool  & percentage of population without a high school diploma 2019\\ 
 & population & state population on 2019 \\
   \hline 
  political   & political  & political engagement score \\
  & republican & percentage prefer republican subtract percentage prefer democrat \\ 
    & swing\_state  & binary score of 0 (not swing state) or 1 (swing state) \\
  \hline
    platform & adoption & adoption rate of Reddit \\
   \hline
\end{tabular}
\end{table}




\vspace{0.15cm}
\noindent
\textbf{Classification of News Comments.}
 The circulation of reputable news significantly outnumbered non-reputable news (Table~\ref{tab:dat}): 
  7.6M (93\%) comments contained reputable news articles, but only 116.2K contained fake news articles. We also observed that reputable news sites attracted, on average, only 36 Reddit comments, low-credibility 26, and fake 8. Those low average values are due to the frequency distribution of the number of  comments per news site being skewed:  most news sites attract a few comments only, while a few attract most comments (e.g., approximately one-fifth of all fake news comments contained URLs from \url{breitbart.com}). To then ascertain that our localization procedure did not select a specific type of user but selected a set representative  of the general user population, we compared the 3M users with assigned locations to another 3M users without locations. We observed that the average numbers of  comments posted by users of the two groups were comparable, with just a small difference:  1.7\% of all geotagged users had posted at least 1 comment containing fake news URLs, whereas only 0.6\% of non-geotagged users did. This difference can be explained by non-geotagged users being less invested in U.S. news as, on average, they are less likely to all be from the U.S.. Additionally, while our lists of news sites are widely popular in researching misinformation, prior work had highlighted that the different lists had being created using  varying  labeling procedures~\cite{bozarth2020higher}. As such, we included additional steps detailed in Supplementary Material to validate our news site classification approach. Briefly, we compared our labels ({\it fake, lowcred}, and {\it reputable}) to trustworthiness scores of news sites provided by professional fact-checkers~\cite{pennycook2019fighting}, and observed that reputable news sites had the highest average trustworthiness score (0.66), followed by low-credibility news sites (0.10), and  finally fake news sites (0.02),  suggesting that our labels were well aligned with the ratings of professional fact-checkers.


\begin{figure}[!t]
    \centering
    \includegraphics[width=0.95\linewidth]{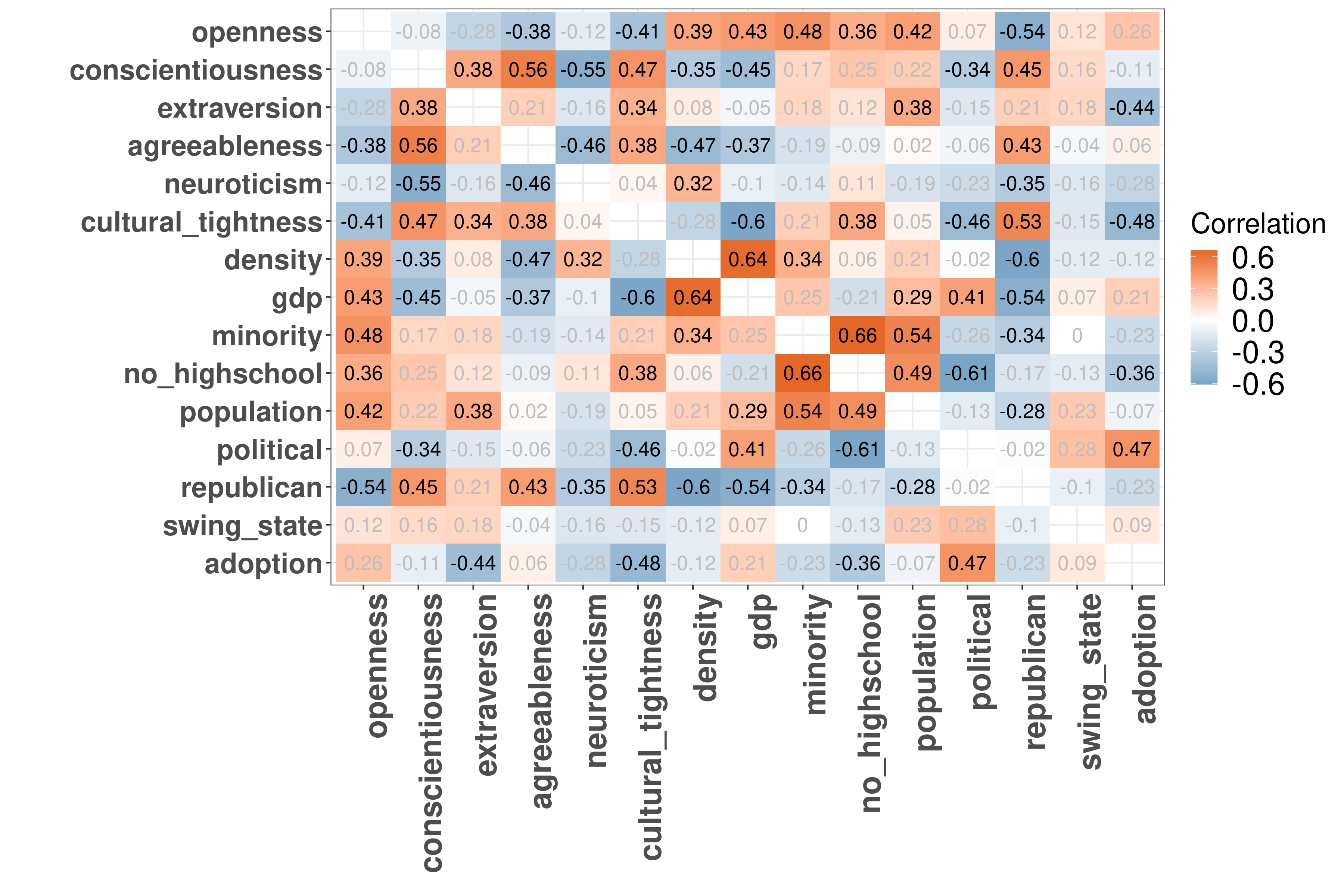}
    \caption{Cross-correlation between State-level Factors. Statistically insignificant correlations ($p$-value$\ge 0.05$) are grayed out.}
    \label{fig:statecorrs}
\end{figure}


\vspace{0.15cm}
{\bf \noindent State-level Attributes.} We included the following state-level attributes that were shown by prior studies to be indicative of individual and community's tendency to share misinformation~\cite{forgas2019social,khan2019recognise,grinberg2019fake}. These attributes were categorized into personality and cultural factors, socio-economic conditions, and political attributes (Table~\ref{tab:stateattr}).

\vspace{0.15cm}
{\it\noindent Personality and Culture:} Prior work had observed significant individual-level associations between personality/culture and circulation of misinformation~\cite{forgas2019social,balestrucci2020credulous,bonney2018fake,islam2020misinformation}. {For instance, individuals scoring high in conscientiousness are significantly less likely to spread false content~\cite{forgas2019social}. Similarly, a lower level of extraversion is associated with a higher discernment of misinformation~\cite{calvillo2021personality}. Therefore, we included the big 5 personality variables of openness, conscientiousness,  extroversion, agreeableness, and neuroticism (OCEAN)~\cite{rentfrow2013divided}, and added cultural tightness to them. This latter variable reflects the propensity of  
holding adherence to norms in high regard~\cite{harrington2014tightness}, and might well be hindering the spreading of misinformation. 


\vspace{0.15cm}
{\it\noindent Socio-economic:} Some socioeconomic factors  are indicative of an individual's political knowledge, information literacy, and tendency to consume and diffuse misinformation~\cite{allcott2018trends,grinberg2019fake,khan2019recognise}. As an example, individuals who are socio-economically well-off  tend to have more political knowledge~\cite{mcleod1994direct}, which is associated with having a better ability in
 telling apart factual news from misinformation~\cite{khan2019recognise}. Overall, in terms of socio-economic indicators, we included five variables available from the 2019 American Community Survey:
 population ({\it population}); population density as a proxy for urbanization ({\it density});  percentage of population over 25 years old without high school diploma ({\it no\_highschool});  percentage of person of color ({\it minority}); and  gdp per capita ({\it gdp}).

\vspace{0.15cm}
{\it\noindent Political:} An extensive literature had shown that fake news is politically driven and is more likely to be consumed and shared by conservative-leaning individuals and online communities~\cite{allcott2018trends,grinberg2019fake,khan2019recognise,guess2019less,jones2021does,he2019online}. Therefore, we postulated that states' political attributes would be the most indicative of the states' tendency to circulate misinformation, an consequently included three political attributes:  percentage gap between the population leaning towards the Republican party and that leaning towards the Democratic party ({\it republican}) provided by the 2016 Gallup Poll;  whether a state was a battleground state during the 2016 presidential election or not ({\it swing\_state}) provided by the Center for Politics; and the political engagement score ({\it political}) from~\cite{mccann_2020}, which was calculated using the weighted sum of multiple metrics (i.e., percentage of registered voters, total political contribution, and percentage of residents who participated in local political) provided between 2016 and 2019 by the American Community Survey, the U.S. Census Bureau, the Center for Responsive Politics, and Ballotpedia.

\vspace{0.15cm}
{\noindent}
To those socio-economic attributes, we added a state's Reddit adoption rate as a control varaible. That is because online news circulation might well be explained by online adoption rates, which, in turn, happened to be  correlated with some of the socio-economic attributes in our case (Figure~\ref{fig:statecorrs}): negatively with {\it extraversion, cultural\_tightness}, and {\it no\_highschool}, and positively with  {\it political}.  In other words, states that are social, culturally restrictive, and have low education attainment have fewer-than-expected users on Reddit.

\section*{Methods}
\label{sec:method}

\subsection*{Scaling laws of news circulation}
To study circulation within states, we resorted to urban science research in the area of complex systems~\cite{west2017scale,bettencourt2007growth}. Such work has shown that a variety of urban measures such as number of patents and income are power-law functions of population size~\cite{bettencourt2007growth,bonaventura21}. Yet, we do not know whether that is the case for news circulation online: critics might rightly say that the process of online circulation may have little to do with a user's offline conditions or may be just ``too complex'' to be subject to laws.  

To investigate the relationship between news circulation and population size, we used a  methodology that was put forth by Bettencourt \emph{et al.}~\cite{bettencourt2007growth}. Say that $Y$ denotes circulation within a state, then this power-law dependency translates into saying that $Y = constant \cdot N^\beta$. By then taking the log of both sides, we obtain: $\log(Y) = \beta\cdot \log(N) + constant$, where $N$ is the population size, $constant$ is a normalization constant, and $\beta$ is the so-called \emph{scaling exponent}. Typically, the values of this scaling exponent are grouped in three ranges:

\begin{description}[noitemsep]
\item $0.8 > \beta$ (\emph{sublinear})  is found for material quantities  displaying \emph{economies of scale} (e.g., infrastructure); 

\item $0.8\leq \beta < 1.1$ (\emph{linear}) is found for  individual human needs (e.g., jobs, houses);

\item $1.1 \leq \beta < 1.3$   (\emph{superlinear}) is found for measures reflecting wealth creation and innovation  with \emph{increasing returns}, which are typically associated with the intrinsically social nature of large cities (e.g., number of patents, number of successful startups).
\end{description}

\subsection*{Three types of news}
Since the number of Reddit users alone could explain a great portion of the variance for the presence of the three types of news, we used the following approach to separate the impact of platform adoption and the characteristics of a state. Given a news type $s \in \{lowcred, fake, reputable\}$ and state $i$, let $\beta_s$ be the scaling exponent for news type $s$, $f_{s,i}$ denote the total number of news items of type $s$ posted by users from $i$ (in log value), and $N_i$ be the number of users in state $i$ (in log value). We then run the simple regression $f_{s, i}= \beta_s N_{i} + \epsilon_{s,i}$ to determine the residual $\epsilon_{s, i}$, which we call the \emph{Circulation(s,i)} score of state $i$ for the news type $s$. This is the portion of the circulation of news of type $s$ in a state $i$ that is not explained by the number of users in $i$.  Next, we took that residual and run the following model:

\begin{equation}
\label{eq:circulation}
    Circulation({s,i}) = \beta_0^\prime + \beta_1^\prime * v_1 + \beta_2^\prime * v_2 ... + \beta_n^\prime * v_n + \epsilon^\prime,
\end{equation}
where $v_1$, $v_2$, and $v_n$ are the predictors listed in Table~\ref{tab:stateattr}. Note that all variables were standardized with $z$-scores to make regression coefficients easier to interpret. For comparability's sake, in addition to this circulation metric based on the residual, we also used the average number of misinformation news comments as as an alternative metric (i.e., $Circulation({s,i})$ was calculated as the average number of comments containing URLs to news type $s$ posted by Reddit users from state $i$), and reported the results in  Supplementary Material; both metrics showed comparable results.


\section*{Results}
\begin{figure}[t]
\begin{subfigure}[t]{0.5\textwidth}
    \includegraphics[width=0.95\linewidth]{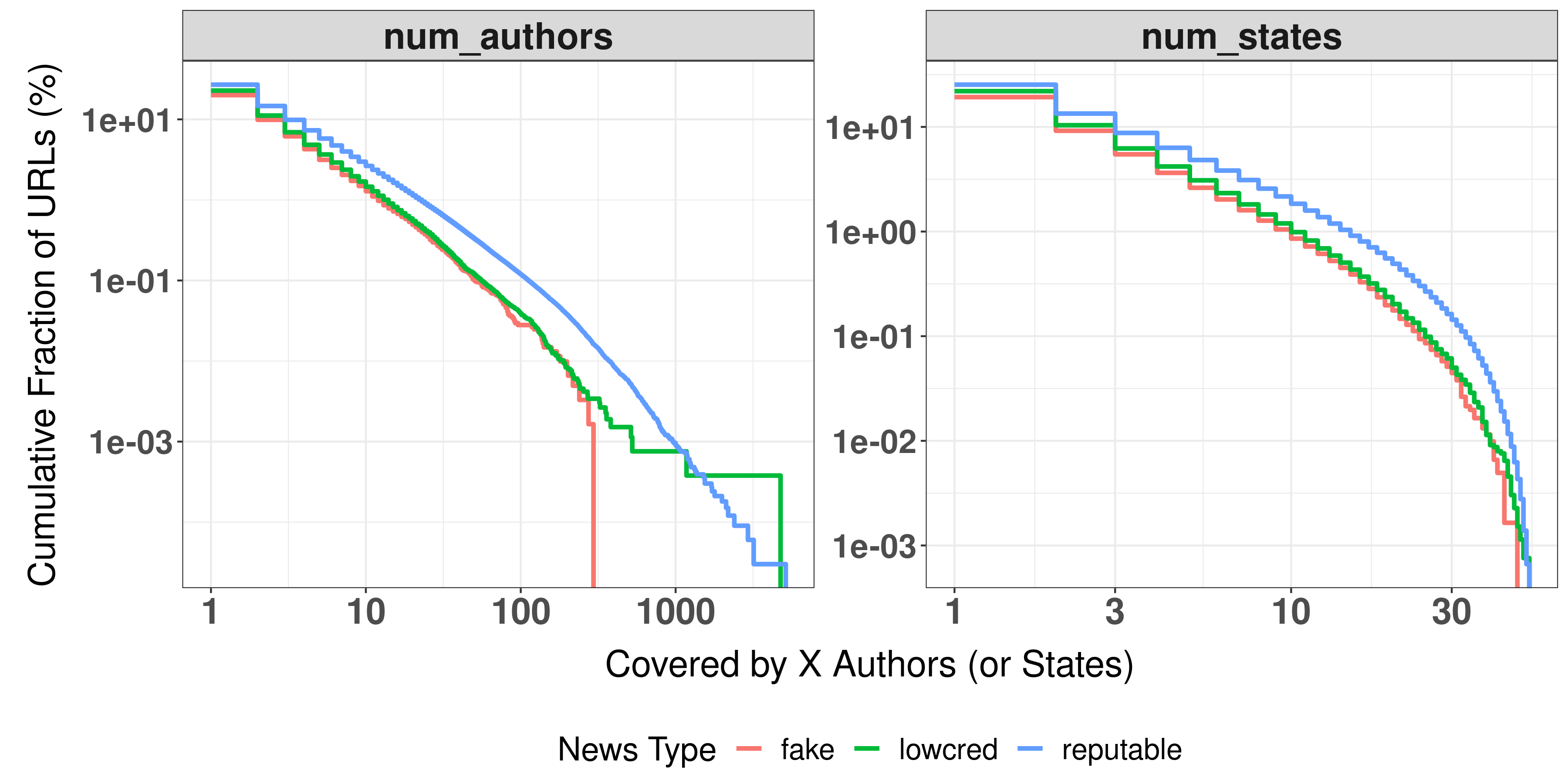}
    \caption{\textbf{Diffusion Reach.} Cumulative fraction of news articles that reached at least a given number ($x$-axis value) of authors or states.\newline We saw that approximately 90\% of all news articles were only\newline posted by 1 or 2 users irrespective of news type.}
    \label{fig:viral1}
\end{subfigure}%
\begin{subfigure}[t]{0.5\textwidth}
    \includegraphics[width=.95\linewidth]{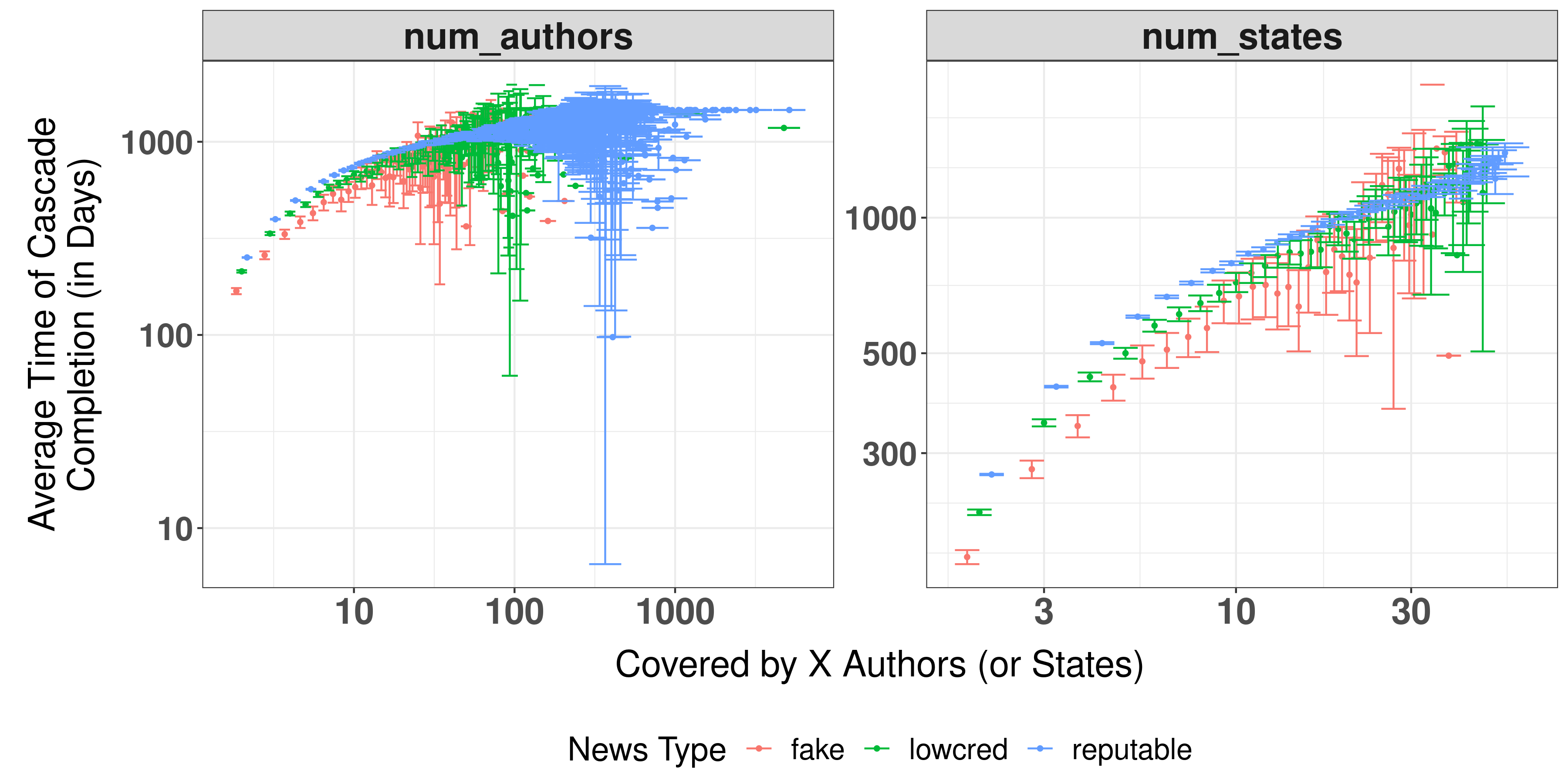}
    \caption{\textbf{Diffusion Speed.} Average cascading time for news articles that reached at least a given number ($x$-axis value) of authors or states. We saw that {\it fake} news tended to take less time on average to reach the same numbers of states and subreddits than {\it reputable} news. However, cascading time was exceedingly long for all news types: for example, average cascading time for news that reached 2 states was 226 days, and for those that reached 5 states was 522 days.}
    \label{fig:viral2}
\end{subfigure}
\label{fig:virals}
\caption{Diffusion for the Three Types of News on Reddit.}
\end{figure}

\subsection*{The role of platform-facilitated news diffusion}
We compared the virality of low-credibility and fake articles to that of reputable news. For each type of news, we generated the cumulative fraction of articles that reached at least a given number of authors or states (Figure~\ref{fig:viral1}). We observed that geographical diffusion is rare on Reddit. More specifically, 74.8\% of all reputable news articles were only posted by a single user who was located in the U.S., and 86.7\% by at most 2 users. The values were comparable for fake and low-credibility news. Additionally, the number of  news URLs that were posted by 5 or more states was only 209.7K for (6.3\% of) reputable news comments,
11.0K for (4.8\% of) low-credibility ones, and 2.23K for (4.2\% of) fake ones. Furthermore, we also observed that the time gaps between the comments were lengthy (Figure~\ref{fig:viral2}). For example, for all news URLs that reached exactly 5 states (only 6\% of news had reached 5 or more states), the average cascading time was over a year. We also ran analysis using the median cascading time, and results were similar. In sum, our results  demonstrate that circulation of news on Reddit is unlikely to be a function of diffusion, and there are several likely explanations for it. First, to reduce content duplication, Reddit moderators typically discourage users from reposting the same content on the same subreddit or even on different subreddits~\cite{richterich2014karma}. Another explanation could be geographical segregation. As the literature showed for platforms other than Reddit~\cite{liben2005geographic,kuchler2021jue}, online users who live far away could be less likely to interact  with each other, thus reducing out-of-state news circulation in the case of Reddit. Our data allowed us to test  this latter explanation, and we did so next.

\subsubsection*{The role of geographical proximity}
To test the extent to which online interactions are impacted by geographical distance, we adopted a metric from related work~\cite{liben2005geographic}. More specifically, we first generated a user-to-user comment network in which an edge exists between a pair of users, if one user had commented on the other's comment/post~\cite{joglekar2020analysing}. The resulting network was unidirectional and weighted.  We then computed the probability of having had an interaction, denoted as $Connectivity_d$, between a pair of users who are at $d$ physical distance apart (measured in km). The distance $d$ between a pair of users was calculated as the distance between the geographical centers of the states that the pair resided in (users from the same state have $d=0$). Mathematically, for a fixed distance $d$ where $d=\{0km, 100km, 200km, 300km...\}$, we calculated $Connectivity_d$ as:

\begin{equation}
    Connectivity_{d} = \frac{|comments_{{i,j}}|_d}{\frac{1}{2}*N_d*(N_d-1)},
\end{equation}

where $N_d$ is the total number of users that were approximately $d$ distance apart offline, and $|comments_{{i,j}}|_d$ is the total number of unique pairs of users who lived $d$ distance apart and who interacted on Reddit (this number is the corresponding weight on the user-to-user comment network). The denominator $\frac{1}{2}*N_d*(N_d-1)$ is the total number of possible user pairs at distance $d$. In other words, given $d$, $Connectivity_d$ is the number of user pairs that interacted with each other normalized by the total number of possible user pairs. We then plotted the logged $Connectivity_d$ in relation to the logged physical distance $d$ in Figure~\ref{fig:geoprob} (red line).  Consistent with prior work~\cite{liben2005geographic}, we found that $Connectivity_d$ rapidly decreases with $d$. For instance, users located approximately 100km apart had $4.35\mathrm{e}{-5}$ probability of interacting with each other via comments. Whereas, the probability decreased to $2.6\mathrm{e}{-5}$ for users located 1000km apart. In other words, geographic proximity increases the probability of interacting (i.e., users located closer in physical distance are more likely to interact with each other): indeed, the probability of interacting is highest for users of the same state ($1.02\mathrm{e}{-4}$) as it is one order of magnitude higher than the out-of-state's probability ( $\ge 2.6\mathrm{e}{-5}$). Next, to ensure that our observation was not primarily driven by  interactions on location-specific subreddits (e.g., r/seattle, r/california), we also limited the {\it scope} of interaction to non-location subreddits. To that end, we updated the definition of $|comments_{{i,j}}|_d$ to be the number of unique pairs of users who lived $d$ distance apart and, crucially, who also had interacted on subreddits that do not have a geographical component. We found that the red and green lines overlap (Figure~\ref{fig:geoprob}), and that 
non-geographically salient users still preferentially interacted with others in closer geographical proximity (green line), suggesting that the observed decay with distance was not dependent on our localization procedure. That is not entirely surprising as 
 online interactions have been shown to be bounded by geography, not least because social networks are based on real-world friends/contacts (as an example, we applied the same $Connectivity_d$ formula to a publicly available Facebook graph, and, in Supplementary Material, we observe that interactions on Facebook  are even more geographically bounded than those on Reddit). Yet, in the case of Reddit, this result is remarkable because the platform is an anonymous forum where both a user's identity and  physical location are  hidden from other users.

\begin{figure}[t]
    \centering
    \includegraphics[width=0.5\linewidth]{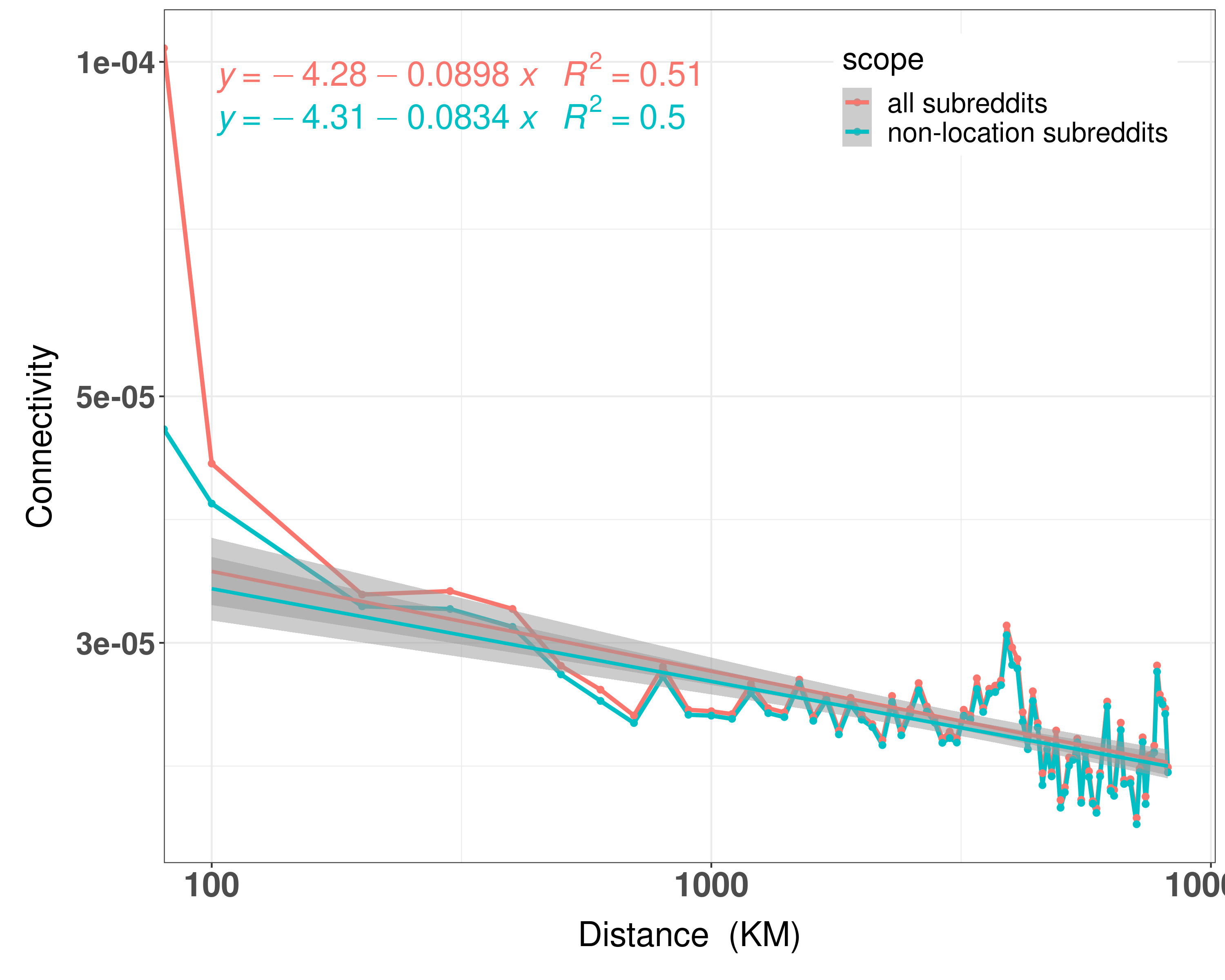}
    \caption{Geographic Distance \emph{vs.} $Connectivity$. The $x$-axis denotes the geographical distance between states' centers and the $y$-axis is the probability that a pair of users with $x$ distance apart offline had interacted with each other on Reddit. Finally, the color denotes the scope of interaction. We surprisingly saw that even for subreddits without an inherent geographical affiliation, users still preferred to interact with others of closer geographical proximity.}
    \label{fig:geoprob}
\end{figure}



\begin{figure}[t]
    \centering
    \includegraphics[width=\linewidth]{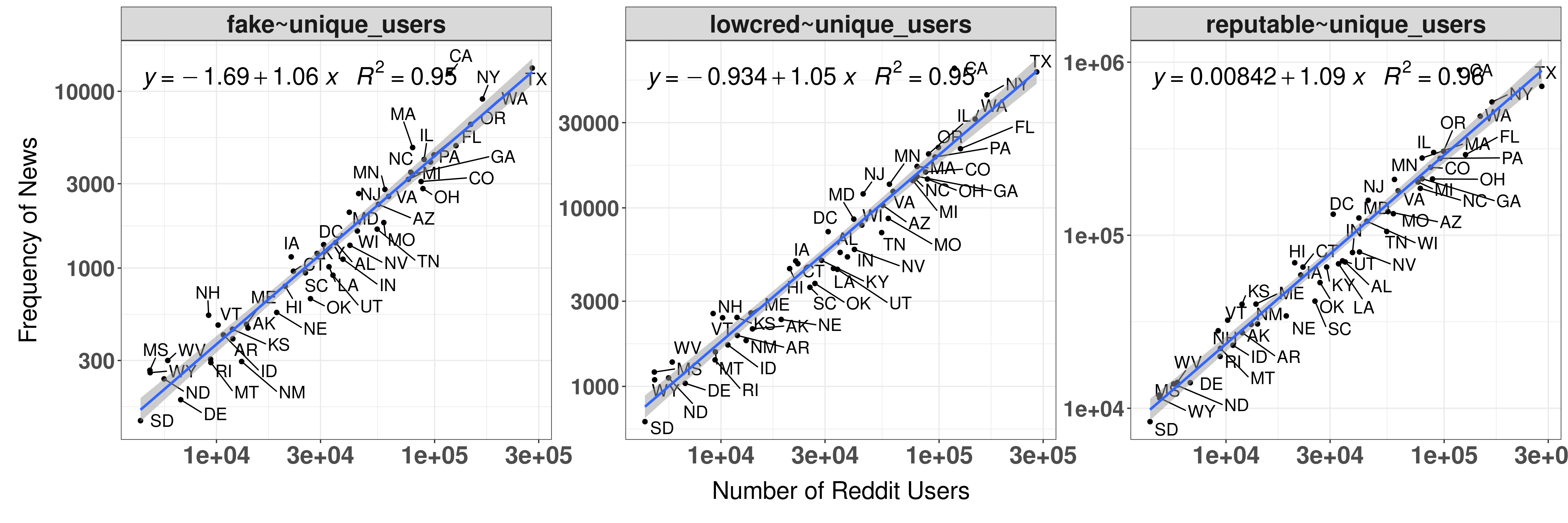}
    \caption{The Scaling of News Circulation. The $x$-axis is the total number of Reddit users from a state, and the $y$-axis denotes the number of posts containing each of the three types of news. We observed that the circulation of news approximates a supply and demand system (i.e., $\beta \approx 1.0$).} 
    \label{fig:residual}
\end{figure}

\subsection*{The scaling laws of news circulation}
Given that interactions are geographically bounded, it was reasonable to hypothesize that a state's news circulation is best explained by the state's variables rather than platform-specific variables. As previously mentioned, based on the scaling laws literature, one of these state variables is  the number of users. As a case in point, consider that the state of California had a total of 60.4K posts containing low-credibility news posts, the highest amongst all the states. It is also the most populous state with the highest number of Reddit users. Similarly, the state of Wyoming had only 1.1K low-credibility posts and is also the least populous with the lowest number of Reddit users. We indeed found evidence that the number of Reddit users in a state is an important predictor of news circulation. It alone explained 95\% ($R\approx 0.95$) of the variance:  1 unit log scale gain in number of users is approximately correlated with exactly 1 unit log scale gain in news circulation ($\beta \approx1$) for all three types of news (Figure~\ref{fig:residual}), suggesting that news circulation on Reddit works as a supply-and-demand system.

\subsection*{The role of the Big Sort}
\begin{figure}[t]
    \centering
    \includegraphics[width=0.8\linewidth]{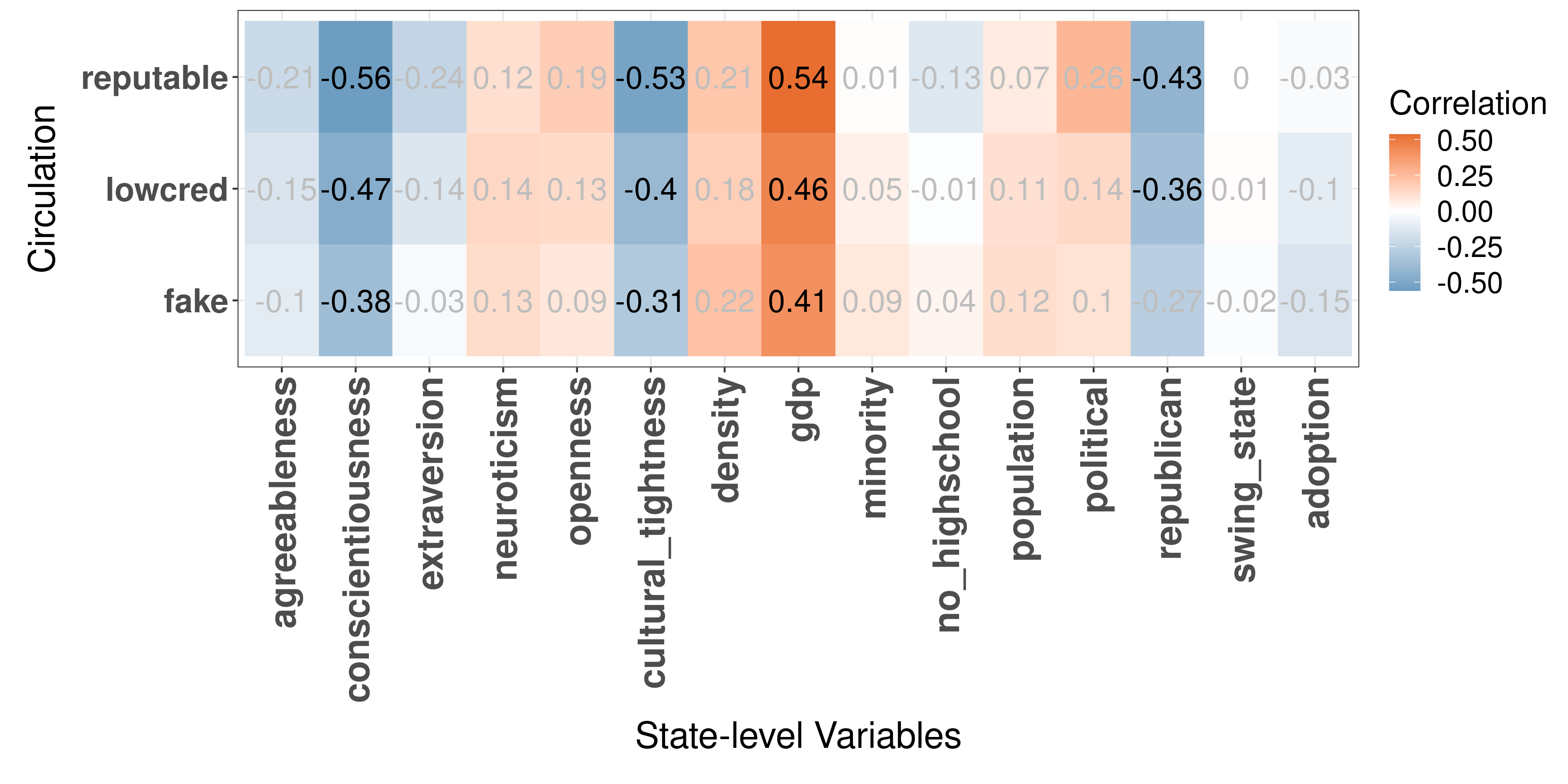}
    \caption{Correlation between Circulation and each independent Variable. Statistically insignificant correlations  ($p$-value$\ge 0.05$) are grayed out.}
    \label{fig:dv_correlation}
\end{figure}

\begin{table}[t] \centering \scriptsize
  \caption{Circulation Regression Results. The {\it personality and culture} models (1)(3)(5) only used personality and cultural explanatory variables. The {\it complete} models (2)(4)(6) used all explanatory variables. For all models, stepAIC  selected the most predictive subset of predictors.} 
  \label{tab:circulation} 
\begin{tabular}{@{\extracolsep{-5pt}}lp{2.2cm} p{2.2cm}  p{2.2cm} p{2.2cm} p{2.2cm} p{2.2cm}} 
\\[-1.8ex]\hline 
\hline \\[-1.8ex] 
 & \multicolumn{6}{c}{\textit{Dependent variable: circulation}} \\ 
\cline{2-7} 
 & {reputable (personality and culture)} & {reputable (complete)} & {lowcred (personality and culture)} & {lowcred (complete)} & {fake (personality and culture)} & {fake (complete)} \\ 
\\[-1.8ex] & {(1)} & {(2)} & {(3)} & {(4)} & {(5)} & {(6)}\\ 
\hline \\[-1.8ex] 
 agreeableness & 0.022 (0.014) & 0.022 (0.015) &  & 0.038$^{**}$ (0.017) &  & 0.033 (0.020) \\ 
  conscientiousness & $-$0.052$^{***}$ (0.015) & $-$0.053$^{***}$ (0.015) & $-$0.040$^{**}$ (0.016) & $-$0.054$^{***}$ (0.018) & $-$0.044$^{***}$ (0.016) & $-$0.040$^{*}$ (0.021) \\ 
  openness &  &  &  &  &  & $-$0.034 (0.023) \\ 
  cultural\_tightness & $-$0.038$^{***}$ (0.014) & $-$0.025 (0.016) & $-$0.025 (0.016) & $-$0.027 (0.018) &  & $-$0.036 (0.024) \\ 
  \hline
  no\_highschool &  &  &  & 0.032$^{**}$ (0.015) &  & 0.054$^{**}$ (0.021) \\ 
  gdp &  & 0.052$^{**}$ (0.019) &  & 0.030$^{*}$ (0.017) &  & 0.046$^{**}$ (0.020) \\ 
  density &  & $-$0.027 (0.017) &  &  &  &  \\ 
    \hline
  political &  & $-$0.023 (0.014) &  &  &  &  \\ 
    \hline
  Constant & $-$0.006 (0.012) & $-$0.006 (0.011) & $-$0.002 (0.014) & $-$0.002 (0.013) & 0.001 (0.016) & 0.001 (0.015) \\ 
 \hline \\[-1.8ex] 
Observations & 48 & 48 & 48 & 48 & 48 & 48 \\ 
R$^{2}$ & 0.432 & 0.521 & 0.261 & 0.401 & 0.142 & 0.349 \\ 
Adjusted R$^{2}$ & 0.393 & 0.451 & 0.228 & 0.329 & 0.123 & 0.254 \\ 
Residual Std. Error & 0.081 (df = 44) & 0.077 (df = 41) & 0.096 (df = 45) & 0.089 (df = 42) & 0.110 (df = 46) & 0.102 (df = 41) \\ 
F Statistic & 11.158$^{***}$ (df = 3; 44) & 7.438$^{***}$ (df = 6; 41) & 7.951$^{***}$ (df = 2; 45) & 5.616$^{***}$ (df = 5; 42) & 7.608$^{***}$ (df = 1; 46) & 3.665$^{***}$ (df = 6; 41) \\ 
\hline 
\hline \\[-1.8ex] 
& \multicolumn{6}{r}{$^{*}$p$<$0.1; $^{**}$p$<$0.05; $^{***}$p$<$0.01} \\ 
\end{tabular} 
\end{table}

\vspace{0.15cm}

To explore why misinformation circulation might deviate from the supply-and-demand model at times, we studied the associations between the news circulation residual metric $Circulation(s, i)$ and state-level attributes. Cultural tightness  and conscientiousness had the highest correlation (absolute value) with circulation across all news types (Figure~\ref{fig:dv_correlation}), not least because the two variables are correlated with each other ($r[cultural\_tightness, conscientiousness]= 0.47, p < 0.05$ in Figure~\ref{fig:statecorrs}). This translates into saying that conscientious states with restrictive social norms circulated fewer news items than what was expected by their Reddit adoption. The association was even more prominent for {\it reputable} news. For example, the correlation between {\it cultural tightness} and $Circulation$ for {\it fake} news was $-0.31$; the correlation was $-0.53$ for {\it reputable} news. In other words, users from states ranked high in {\it conscientiousness} were posting fewer reputable and fake news items than what was expected from their numbers of Reddit users. Next, focusing on political variables, we found that the presence of {\it republican} voters was noticeably negatively correlated with circulation of reputable and low-credibility news but not of fake news (in Figure~\ref{fig:dv_correlation}, $r[circulation,republican]$ is negative for $reputable$ and $lowcred$, but becomes insignificant for $fake$). That result is in line with prior studies showing  that the  majority of misinformation is conservative-leaning~\cite{grinberg2019fake,calvillo2020political}. Also, that result has an additional explanation:  states that  are slightly more likely to use Reddit are democratic ones ($r[adoption, republican]= - 0.23, p >= 0.05$ in Figure~\ref{fig:statecorrs}), as further detailed in Supplementary Material. Surprisingly, we also saw that swing states with competitive political races were not more likely to circulate significantly more news. Finally, focusing on socioeconomic factors, we observed that wealthy states had higher circulation, irrespective of news types.

Next, we focused on the combined effects of state-level attributes by studying each news type separately. For each, we ran 3 partial regression models (personality and culture, socio-economic, and political) plus one combined model. Each of the models (3 partial + 1 complete) was then fitted using {\it stepAIC}, a method that statistically identifies the best combination of independent variables with the lowest AIC value (lower AIC values indicate a better-fit model).  Since we are interested in  
which class of variables (personality and culture \emph{vs.} socio-economic \emph{vs.} political) best explained misinformation circulation, we only report the complete model and the personality and culture partial model here, and report the other 2 partial models  in Supplementary Material. We found that the complete models were able to explain a considerable fraction of variances in circulation residual (adjusted $R^2\approx \{0.25, 0.45\}$ in Table~\ref{tab:circulation}). Further, the variable {\it conscientiousness} was a significant indicator for lower-than-expected circulation for all types of news for all models; whereas {\it gdp} was significantly correlated with higher-than-expected circulation. More interestingly, we also saw that, for the personality and culture partial models, the adjusted $R^2 \approx \{0.12, 0.39\}$. In other words, the $R^2$ differences between the personality and culture models and the complete models were small. As an example, the adjusted $R^2$ for the full model for reputable news was 0.45, whereas the adjusted $R^2$ for the personality and culture model was 0.39 (a difference of only 0.06). In fact, including personality and cultural variables improved the full models' adjusted $R^2$ from 0.10 to 0.20 (see Supplementary Material). Additionally, we also saw that personality and culture models had higher adjusted $R^2$ values than, as Supplementary Material shows, models that exclusively used socioeconomic conditions (adjusted $R^2\approx \{0.15, 0.29\}$) or political characteristics (adjusted $R^2\approx \{0.06, 0.21\}$). As a robustness check,  we also reran our analysis using normalized circulation volume. Specifically, we redefined $Circulation(s, i)$ as the average number of comments containing URLs to news type $s$ posted by Reddit users from state $i$. We then reran Equation~(\ref{eq:circulation}). The main findings detailed in Supplementary Material did not change: personality and cultural factors still remained strong indicators of circulation.

Finally, by comparing the values of the beta coefficients for different news types in Table~\ref{tab:circulation}, we observed that circulation of any news types was facilitated in states that: are wealthier ($gdp$ has positive $beta$'s in Table~\ref{tab:circulation}), have residents who are less diligent in terms of personality ($conscientiousness$ has negative $beta$'s), and are characterized by loose cultures which understate the importance of adherence to norms ($cultural\_tightness$ has negative $beta$'s). That holds for all types of news. We then focused on the circulation of misinformation specifically, and observed that was taking place once these three factors were combined with a fourth one: low education levels ($no\_highschool$ has a positive  $beta$  in the complete fake news model in Table~\ref{tab:circulation}).

 \section*{Discussion}

We have focused on examining factors associated with the geographical circulation of both reputable, fake and low credibility news on Reddit. We observed that users' online interactions were mediated by their offline physical proximity, with users located far away from each other having a significantly lower probability of interacting, despite their real identities and locations not being visible. Further, news content was rarely circulated across states, likely the result of both  interactions being geographically constrained and  Reddit's platform guidelines (e.g., Reddit discourages content reposts~\cite{jhaver2019does}). Overall, we found that an individual user's news consumption was influenced by their state's characteristics rather than by some contagion effects on the platform. Interestingly, these characteristics had  more to do with cultural values than with political beliefs. 

Our work has two main ramifications for research focused on ``why'' do people share fake news and on ``how'' to curtail the spread of misinformation. The first has to do with the stability of personality and culture. More specifically, personality and culture are ingrained parts of every individual; they generally remain stable for people who have reached adulthood~\cite{mccrae1994stability}. Moreover, past research showed that individuals are likely drawn to regions that match their personality and cultural norms as this matching increases their overall life satisfaction~\cite{jokela2015geographically}. In fact, prior longitudinal analysis on state-wide personality traits showed that states' big-5 personality ranks remained unchanged in the last 20 years~\cite{elleman2018personality}. Given such level of ``stability'' and clustering, these traits are likely to make combating misinformation more difficult  (for instance, it would be difficult to compel ``unconscientious personalities'' to be more conscientious~\cite{schurer2015universities}). 

The second ramification is focused on the role of social media. Social media platforms' recommendation and personalization algorithms had led to the formulation of homogeneous, tight-knit communities en mass. These communities had then facilitated the circulation of misinformation. Thus, researchers had proposed various ways to regulate these algorithms, including increasing the diversity of perspectives and connections available to users. Yet, our results  suggest that algorithmic amplification is  not the root cause of misinformation, at least  not in the case of Reddit. Rather, the root cause has to be found in the   geographic sorting that has been happening in the last 40 years. Given these considerations, we argue that a more productive way to combat misinformation is to reduce its production altogether. That is, we need to disincentivize the creation of fake and low-credibility news sites and news content before they can be shared by individuals and online communities. This can be done in several ways. For instance, many fake news sites are driven by ad profit~\cite{bakir2018fake}. As such, ad firms and retailers can curtail misinformation by blacklisting known fake and low-credibility news sites, and recent research suggested that, in so doing, major ad firms would not suffer any significant loss of revenues~\cite{bozarth2021market}. Similarly, lawmakers can also pass regulations such as criminalizing false stories (e.g., laws against defamation in the offline world already exist) with the potential to ignite communal tension~\cite{feingold2017fake}. More generally, our results speak to the importance of addressing questions relating to social justice. Issues subject of misinformation are complex phenomena, and may include questions of societal trust in institutions. Marginalized communities tend to cultivate mistrust because they have been negatively affected by the actions of those institutions~\cite{royal2022online} and, as such, trust between these communities and public institutions has to be rebuilt. 



There are four main limitations to our work. First, our work was exclusively focused on circulation  of misinformation, and, as such, we did not address its actual consumption (e.g., we cannot determine the number of users who actually read and believed the content from the posted news URLs, but could only determine the number of those who were potentially exposed).

Second, our project solely relied on Reddit data, and we do not know whether our results generalize to other platforms. Yet, in Supplementary Material, we showed that interactions on Facebook are even more geographically localized that those on Reddit, suggesting that geographic segregation might play an even stronger role on Facebook. 

Third, we approximated a user's geolocation at the state level because that was the granularity allowed by Reddit. However, a state's personality and culture, socioeconomic and political attributes can vary significantly from one sub-region to another. For instance, many prior studies had identified significant differences between rural and urban areas of the U.S.~\cite{hindman2000rural}. Future work on platforms other than Reddit should consider using more granular location data whenever possible. 

Fourth, our data  did not contain comments that were deleted prior to being collected by~\url{pushshift.io}. As such, we could not examine whether those deleted comments contained news URLs. In particular, comments that were removed by Automoderator (bots) were unavailable to us, as these comments were removed as soon as they were posted. Nevertheless, the Reddit dataset from ~\url{pushshift.io} remains one of the most comprehensive datasets available~\cite{baumgartner2020pushshift}. Furthermore, reputable news is unlikely to be removed by moderators, and our observations for true news still showed the prominent role of regional personality and culture, speaking to the robustness of our findings.



\bibliography{main}

\begin{thebibliography}{10}
\urlstyle{rm}
\expandafter\ifx\csname url\endcsname\relax
  \def\url#1{\texttt{#1}}\fi
\expandafter\ifx\csname urlprefix\endcsname\relax\def\urlprefix{URL }\fi
\expandafter\ifx\csname doiprefix\endcsname\relax\def\doiprefix{DOI: }\fi
\providecommand{\bibinfo}[2]{#2}
\providecommand{\eprint}[2][]{\url{#2}}

\bibitem{gradon2020crime}
\bibinfo{author}{Grado{\'n}, K.}
\newblock \bibinfo{journal}{\bibinfo{title}{Crime in the time of the plague:
  Fake news pandemic and the challenges to law-enforcement and intelligence
  community}}.
\newblock {\emph{\JournalTitle{Society Register}}}
  \textbf{\bibinfo{volume}{4}}, \bibinfo{pages}{133--148}
  (\bibinfo{year}{2020}).

\bibitem{duong2021covid}
\bibinfo{author}{Duong, H.}
\newblock \bibinfo{journal}{\bibinfo{title}{Covid-19 fake news and attitudes
  toward asian americans}}.
\newblock {\emph{\JournalTitle{Journal of Media Research}}}
  \textbf{\bibinfo{volume}{14}}, \bibinfo{pages}{5--29} (\bibinfo{year}{2021}).

\bibitem{forgas2019social}
\bibinfo{author}{Forgas, J.~P.} \& \bibinfo{author}{Baumeister, R.}
\newblock \emph{\bibinfo{title}{The Social Psychology of Gullibility:
  Conspiracy Theories, Fake News and Irrational Beliefs}}
  (\bibinfo{publisher}{Routledge}, \bibinfo{year}{2019}).

\bibitem{burbach2019shares}
\bibinfo{author}{Burbach, L.}, \bibinfo{author}{Halbach, P.},
  \bibinfo{author}{Ziefle, M.} \& \bibinfo{author}{Calero~Valdez, A.}
\newblock \bibinfo{title}{Who shares fake news in online social networks?}
\newblock In \emph{\bibinfo{booktitle}{Proceedings of the 27th ACM Conference
  on User Modeling, Adaptation and Personalization}}, \bibinfo{pages}{234--242}
  (\bibinfo{year}{2019}).

\bibitem{buchanan2019spreading}
\bibinfo{author}{Buchanan, T.} \& \bibinfo{author}{Benson, V.}
\newblock \bibinfo{journal}{\bibinfo{title}{Spreading disinformation on
  facebook: Do trust in message source, risk propensity, or personality affect
  the organic reach of “fake news”?}}
\newblock {\emph{\JournalTitle{Social Media+ Society}}}
  \textbf{\bibinfo{volume}{5}}, \bibinfo{pages}{2056305119888654}
  (\bibinfo{year}{2019}).

\bibitem{grinberg2019fake}
\bibinfo{author}{Grinberg, N.}, \bibinfo{author}{Joseph, K.},
  \bibinfo{author}{Friedland, L.}, \bibinfo{author}{Swire-Thompson, B.} \&
  \bibinfo{author}{Lazer, D.}
\newblock \bibinfo{journal}{\bibinfo{title}{Fake news on twitter during the
  2016 us presidential election}}.
\newblock {\emph{\JournalTitle{Science}}} \textbf{\bibinfo{volume}{363}},
  \bibinfo{pages}{374--378} (\bibinfo{year}{2019}).

\bibitem{balestrucci2020credulous}
\bibinfo{author}{Balestrucci, A.} \& \bibinfo{author}{De~Nicola, R.}
\newblock \bibinfo{title}{Credulous users and fake news: a real case study on
  the propagation in twitter}.
\newblock In \emph{\bibinfo{booktitle}{2020 IEEE Conference on Evolving and
  Adaptive Intelligent Systems (EAIS)}}, \bibinfo{pages}{1--8}
  (\bibinfo{organization}{IEEE}, \bibinfo{year}{2020}).

\bibitem{scherer2021susceptible}
\bibinfo{author}{Scherer, L.~D.} \emph{et~al.}
\newblock \bibinfo{journal}{\bibinfo{title}{Who is susceptible to online health
  misinformation? a test of four psychosocial hypotheses.}}
\newblock {\emph{\JournalTitle{Health Psychology}}}  (\bibinfo{year}{2021}).

\bibitem{osmundsen2021partisan}
\bibinfo{author}{Osmundsen, M.}, \bibinfo{author}{Bor, A.},
  \bibinfo{author}{Vahlstrup, P.~B.}, \bibinfo{author}{Bechmann, A.} \&
  \bibinfo{author}{Petersen, M.~B.}
\newblock \bibinfo{journal}{\bibinfo{title}{Partisan polarization is the
  primary psychological motivation behind political fake news sharing on
  twitter}}.
\newblock {\emph{\JournalTitle{American Political Science Review}}}
  \bibinfo{pages}{1--17} (\bibinfo{year}{2021}).

\bibitem{aral2020hype}
\bibinfo{author}{Aral, S.}
\newblock \emph{\bibinfo{title}{The Hype Machine: How Social Media Disrupts Our
  Elections, Our Economy, and Our Health--and How We Must Adapt}}
  (\bibinfo{publisher}{Currency}, \bibinfo{year}{2020}).

\bibitem{vosoughi2018spread}
\bibinfo{author}{Vosoughi, S.}, \bibinfo{author}{Roy, D.} \&
  \bibinfo{author}{Aral, S.}
\newblock \bibinfo{journal}{\bibinfo{title}{The spread of true and false news
  online}}.
\newblock {\emph{\JournalTitle{Science}}} \textbf{\bibinfo{volume}{359}},
  \bibinfo{pages}{1146--1151}, \doiprefix\url{10.1126/science.aap9559}
  (\bibinfo{year}{2018}).

\bibitem{cinelli2021echo}
\bibinfo{author}{Cinelli, M.}, \bibinfo{author}{Morales, G. D.~F.},
  \bibinfo{author}{Galeazzi, A.}, \bibinfo{author}{Quattrociocchi, W.} \&
  \bibinfo{author}{Starnini, M.}
\newblock \bibinfo{journal}{\bibinfo{title}{The echo chamber effect on social
  media}}.
\newblock {\emph{\JournalTitle{Proceedings of the National Academy of
  Sciences}}} \textbf{\bibinfo{volume}{118}} (\bibinfo{year}{2021}).

\bibitem{guess2021cracking}
\bibinfo{author}{Guess, A.}, \bibinfo{author}{Aslett, K.},
  \bibinfo{author}{Tucker, J.}, \bibinfo{author}{Bonneau, R.} \&
  \bibinfo{author}{Nagler, J.}
\newblock \bibinfo{journal}{\bibinfo{title}{Cracking open the news feed:
  Exploring what us facebook users see and share with large-scale platform
  data}}.
\newblock {\emph{\JournalTitle{Journal of Quantitative Description: Digital
  Media}}} \textbf{\bibinfo{volume}{1}} (\bibinfo{year}{2021}).

\bibitem{pariser2011filter}
\bibinfo{author}{Pariser, E.}
\newblock \emph{\bibinfo{title}{The filter bubble: How the new personalized web
  is changing what we read and how we think}} (\bibinfo{publisher}{Penguin},
  \bibinfo{year}{2011}).

\bibitem{jamieson2008echo}
\bibinfo{author}{Jamieson, K.~H.} \& \bibinfo{author}{Cappella, J.~N.}
\newblock \emph{\bibinfo{title}{Echo chamber: Rush Limbaugh and the
  conservative media establishment}} (\bibinfo{publisher}{Oxford University
  Press}, \bibinfo{year}{2008}).

\bibitem{bakir2018fake}
\bibinfo{author}{Bakir, V.} \& \bibinfo{author}{McStay, A.}
\newblock \bibinfo{journal}{\bibinfo{title}{Fake news and the economy of
  emotions: Problems, causes, solutions}}.
\newblock {\emph{\JournalTitle{Digital journalism}}}
  \textbf{\bibinfo{volume}{6}}, \bibinfo{pages}{154--175}
  (\bibinfo{year}{2018}).

\bibitem{rathje2021out}
\bibinfo{author}{Rathje, S.}, \bibinfo{author}{Van~Bavel, J.~J.} \&
  \bibinfo{author}{van~der Linden, S.}
\newblock \bibinfo{journal}{\bibinfo{title}{Out-group animosity drives
  engagement on social media}}.
\newblock {\emph{\JournalTitle{Proceedings of the National Academy of
  Sciences}}} \textbf{\bibinfo{volume}{118}} (\bibinfo{year}{2021}).

\bibitem{bishop2009big}
\bibinfo{author}{Bishop, B.}
\newblock \emph{\bibinfo{title}{The big sort: Why the clustering of like-minded
  America is tearing us apart}} (\bibinfo{publisher}{Houghton Mifflin
  Harcourt}, \bibinfo{year}{2009}).

\bibitem{glass2014red}
\bibinfo{author}{Glass, J.} \& \bibinfo{author}{Levchak, P.}
\newblock \bibinfo{journal}{\bibinfo{title}{Red states, blue states, and
  divorce: Understanding the impact of conservative protestantism on regional
  variation in divorce rates}}.
\newblock {\emph{\JournalTitle{American Journal of Sociology}}}
  \textbf{\bibinfo{volume}{119}}, \bibinfo{pages}{1002--1046}
  (\bibinfo{year}{2014}).

\bibitem{monson2011all}
\bibinfo{author}{Monson, R.~A.} \& \bibinfo{author}{Mertens, J.~B.}
\newblock \bibinfo{journal}{\bibinfo{title}{All in the family: Red states, blue
  states, and postmodern family patterns, 2000 and 2004}}.
\newblock {\emph{\JournalTitle{The Sociological Quarterly}}}
  \textbf{\bibinfo{volume}{52}}, \bibinfo{pages}{244--267}
  (\bibinfo{year}{2011}).

\bibitem{jokela2015geographically}
\bibinfo{author}{Jokela, M.}, \bibinfo{author}{Bleidorn, W.},
  \bibinfo{author}{Lamb, M.~E.}, \bibinfo{author}{Gosling, S.~D.} \&
  \bibinfo{author}{Rentfrow, P.~J.}
\newblock \bibinfo{journal}{\bibinfo{title}{Geographically varying associations
  between personality and life satisfaction in the london metropolitan area}}.
\newblock {\emph{\JournalTitle{Proceedings of the National Academy of
  Sciences}}} \textbf{\bibinfo{volume}{112}}, \bibinfo{pages}{725--730}
  (\bibinfo{year}{2015}).

\bibitem{scala2017political}
\bibinfo{author}{Scala, D.~J.} \& \bibinfo{author}{Johnson, K.~M.}
\newblock \bibinfo{journal}{\bibinfo{title}{Political polarization along the
  rural-urban continuum? the geography of the presidential vote, 2000--2016}}.
\newblock {\emph{\JournalTitle{The ANNALS of the American Academy of Political
  and Social Science}}} \textbf{\bibinfo{volume}{672}},
  \bibinfo{pages}{162--184} (\bibinfo{year}{2017}).

\bibitem{rentfrow2009statewide}
\bibinfo{author}{Rentfrow, P.~J.}, \bibinfo{author}{Jost, J.~T.},
  \bibinfo{author}{Gosling, S.~D.} \& \bibinfo{author}{Potter, J.}
\newblock \bibinfo{journal}{\bibinfo{title}{Statewide differences in
  personality predict voting patterns in 1996--2004 us presidential
  elections}}.
\newblock {\emph{\JournalTitle{Social and psychological bases of ideology and
  system justification}}} \textbf{\bibinfo{volume}{1}},
  \bibinfo{pages}{314--349} (\bibinfo{year}{2009}).

\bibitem{rentfrow2013divided}
\bibinfo{author}{Rentfrow, P.~J.} \emph{et~al.}
\newblock \bibinfo{journal}{\bibinfo{title}{Divided we stand: Three
  psychological regions of the united states and their political, economic,
  social, and health correlates.}}
\newblock {\emph{\JournalTitle{Journal of personality and social psychology}}}
  \textbf{\bibinfo{volume}{105}}, \bibinfo{pages}{996} (\bibinfo{year}{2013}).

\bibitem{elleman2018personality}
\bibinfo{author}{Elleman, L.~G.}, \bibinfo{author}{Condon, D.~M.},
  \bibinfo{author}{Russin, S.~E.} \& \bibinfo{author}{Revelle, W.}
\newblock \bibinfo{journal}{\bibinfo{title}{The personality of us states:
  Stability from 1999 to 2015}}.
\newblock {\emph{\JournalTitle{Journal of Research in Personality}}}
  \textbf{\bibinfo{volume}{72}}, \bibinfo{pages}{64--72}
  (\bibinfo{year}{2018}).

\bibitem{zannettou2017web}
\bibinfo{author}{Zannettou, S.} \emph{et~al.}
\newblock \bibinfo{title}{The web centipede: understanding how web communities
  influence each other through the lens of mainstream and alternative news
  sources}.
\newblock In \emph{\bibinfo{booktitle}{Proceedings of the 2017 internet
  measurement conference}}, \bibinfo{pages}{405--417} (\bibinfo{year}{2017}).

\bibitem{zannettou2018origins}
\bibinfo{author}{Zannettou, S.} \emph{et~al.}
\newblock \bibinfo{title}{On the origins of memes by means of fringe web
  communities}.
\newblock In \emph{\bibinfo{booktitle}{Proceedings of the Internet Measurement
  Conference 2018}}, \bibinfo{pages}{188--202} (\bibinfo{year}{2018}).

\bibitem{baumgartner2020pushshift}
\bibinfo{author}{Baumgartner, J.}, \bibinfo{author}{Zannettou, S.},
  \bibinfo{author}{Keegan, B.}, \bibinfo{author}{Squire, M.} \&
  \bibinfo{author}{Blackburn, J.}
\newblock \bibinfo{title}{The pushshift reddit dataset}.
\newblock In \emph{\bibinfo{booktitle}{Proceedings of the international AAAI
  conference on web and social media}}, vol.~\bibinfo{volume}{14},
  \bibinfo{pages}{830--839} (\bibinfo{year}{2020}).

\bibitem{balsamo2019firsthand}
\bibinfo{author}{Balsamo, D.}, \bibinfo{author}{Bajardi, P.} \&
  \bibinfo{author}{Panisson, A.}
\newblock \bibinfo{title}{Firsthand opiates abuse on social media: monitoring
  geospatial patterns of interest through a digital cohort}.
\newblock In \emph{\bibinfo{booktitle}{The World Wide Web Conference}},
  \bibinfo{pages}{2572--2579} (\bibinfo{year}{2019}).

\bibitem{bozarth2020higher}
\bibinfo{author}{Bozarth, L.}, \bibinfo{author}{Saraf, A.} \&
  \bibinfo{author}{Budak, C.}
\newblock \bibinfo{title}{Higher ground? how groundtruth labeling impacts our
  understanding of fake news about the 2016 us presidential nominees}.
\newblock In \emph{\bibinfo{booktitle}{Proceedings of the International AAAI
  Conference on Web and Social Media}}, vol.~\bibinfo{volume}{14},
  \bibinfo{pages}{48--59} (\bibinfo{year}{2020}).

\bibitem{vargo2018agenda}
\bibinfo{author}{Vargo, C.~J.}, \bibinfo{author}{Guo, L.} \&
  \bibinfo{author}{Amazeen, M.~A.}
\newblock \bibinfo{journal}{\bibinfo{title}{The agenda-setting power of fake
  news}}.
\newblock {\emph{\JournalTitle{new media \& society}}}
  \textbf{\bibinfo{volume}{20}}, \bibinfo{pages}{2028--2049}
  (\bibinfo{year}{2018}).

\bibitem{zimdars2016my}
\bibinfo{author}{Zimdars, M.}
\newblock \bibinfo{journal}{\bibinfo{title}{My “fake news list” went viral.
  but made-up stories are only part of the problem}}.
\newblock {\emph{\JournalTitle{The Washington Post}}}  (\bibinfo{year}{2016}).

\bibitem{politifact}
\bibinfo{author}{{Politifact staff}}.
\newblock \bibinfo{title}{Politifact guide to fake news websites and what they
  peddle} (\bibinfo{year}{2018}).

\bibitem{dailydot}
\bibinfo{author}{Couts, A.} \& \bibinfo{author}{Wyrich, A.}
\newblock \bibinfo{title}{Here are all the ‘fake news’ sites to watch out
  for on facebook} (\bibinfo{year}{2016}).

\bibitem{allcott2018trends}
\bibinfo{author}{Allcott, H.}, \bibinfo{author}{Gentzkow, M.} \&
  \bibinfo{author}{Yu, C.}
\newblock \bibinfo{journal}{\bibinfo{title}{Trends in the diffusion of
  misinformation on social media}}.
\newblock {\emph{\JournalTitle{arXiv preprint arXiv:1809.05901}}}
  (\bibinfo{year}{2018}).

\bibitem{pennycook2019fighting}
\bibinfo{author}{Pennycook, G.} \& \bibinfo{author}{Rand, D.~G.}
\newblock \bibinfo{journal}{\bibinfo{title}{Fighting misinformation on social
  media using crowdsourced judgments of news source quality}}.
\newblock {\emph{\JournalTitle{Proceedings of the National Academy of
  Sciences}}} \textbf{\bibinfo{volume}{116}}, \bibinfo{pages}{2521--2526}
  (\bibinfo{year}{2019}).

\bibitem{khan2019recognise}
\bibinfo{author}{Khan, M.~L.} \& \bibinfo{author}{Idris, I.~K.}
\newblock \bibinfo{journal}{\bibinfo{title}{Recognise misinformation and verify
  before sharing: a reasoned action and information literacy perspective}}.
\newblock {\emph{\JournalTitle{Behaviour \& Information Technology}}}
  \textbf{\bibinfo{volume}{38}}, \bibinfo{pages}{1194--1212}
  (\bibinfo{year}{2019}).

\bibitem{bonney2018fake}
\bibinfo{author}{Bonney, K.~M.}
\newblock \bibinfo{journal}{\bibinfo{title}{Fake news with real consequences:
  the effect of cultural identity on the perception of science}}.
\newblock {\emph{\JournalTitle{The American Biology Teacher}}}
  \textbf{\bibinfo{volume}{80}}, \bibinfo{pages}{686--688}
  (\bibinfo{year}{2018}).

\bibitem{islam2020misinformation}
\bibinfo{author}{Islam, A.~N.}, \bibinfo{author}{Laato, S.},
  \bibinfo{author}{Talukder, S.} \& \bibinfo{author}{Sutinen, E.}
\newblock \bibinfo{journal}{\bibinfo{title}{Misinformation sharing and social
  media fatigue during covid-19: An affordance and cognitive load
  perspective}}.
\newblock {\emph{\JournalTitle{Technological Forecasting and Social Change}}}
  \textbf{\bibinfo{volume}{159}}, \bibinfo{pages}{120201}
  (\bibinfo{year}{2020}).

\bibitem{calvillo2021personality}
\bibinfo{author}{Calvillo, D.~P.}, \bibinfo{author}{Garcia, R.~J.},
  \bibinfo{author}{Bertrand, K.} \& \bibinfo{author}{Mayers, T.~A.}
\newblock \bibinfo{journal}{\bibinfo{title}{Personality factors and
  self-reported political news consumption predict susceptibility to political
  fake news}}.
\newblock {\emph{\JournalTitle{Personality and Individual Differences}}}
  \textbf{\bibinfo{volume}{174}}, \bibinfo{pages}{110666}
  (\bibinfo{year}{2021}).

\bibitem{harrington2014tightness}
\bibinfo{author}{Harrington, J.~R.} \& \bibinfo{author}{Gelfand, M.~J.}
\newblock \bibinfo{journal}{\bibinfo{title}{Tightness--looseness across the 50
  united states}}.
\newblock {\emph{\JournalTitle{Proceedings of the National Academy of
  Sciences}}} \textbf{\bibinfo{volume}{111}}, \bibinfo{pages}{7990--7995}
  (\bibinfo{year}{2014}).

\bibitem{mcleod1994direct}
\bibinfo{author}{McLeod, D.~M.} \& \bibinfo{author}{Perse, E.~M.}
\newblock \bibinfo{journal}{\bibinfo{title}{Direct and indirect effects of
  socioeconomic status on public affairs knowledge}}.
\newblock {\emph{\JournalTitle{Journalism Quarterly}}}
  \textbf{\bibinfo{volume}{71}}, \bibinfo{pages}{433--442}
  (\bibinfo{year}{1994}).

\bibitem{guess2019less}
\bibinfo{author}{Guess, A.}, \bibinfo{author}{Nagler, J.} \&
  \bibinfo{author}{Tucker, J.}
\newblock \bibinfo{journal}{\bibinfo{title}{Less than you think: Prevalence and
  predictors of fake news dissemination on facebook}}.
\newblock {\emph{\JournalTitle{Science advances}}}
  \textbf{\bibinfo{volume}{5}}, \bibinfo{pages}{eaau4586}
  (\bibinfo{year}{2019}).

\bibitem{jones2021does}
\bibinfo{author}{Jones-Jang, S.~M.}, \bibinfo{author}{Mortensen, T.} \&
  \bibinfo{author}{Liu, J.}
\newblock \bibinfo{journal}{\bibinfo{title}{Does media literacy help
  identification of fake news? information literacy helps, but other literacies
  don’t}}.
\newblock {\emph{\JournalTitle{American Behavioral Scientist}}}
  \textbf{\bibinfo{volume}{65}}, \bibinfo{pages}{371--388}
  (\bibinfo{year}{2021}).

\bibitem{he2019online}
\bibinfo{author}{He, L.}, \bibinfo{author}{Yang, H.}, \bibinfo{author}{Xiong,
  X.} \& \bibinfo{author}{Lai, K.}
\newblock \bibinfo{journal}{\bibinfo{title}{Online rumor transmission among
  younger and older adults}}.
\newblock {\emph{\JournalTitle{Sage open}}} \textbf{\bibinfo{volume}{9}},
  \bibinfo{pages}{2158244019876273} (\bibinfo{year}{2019}).

\bibitem{mccann_2020}
\bibinfo{author}{McCann, A.}
\newblock \bibinfo{title}{Most and least politically engaged states}
  (\bibinfo{year}{2020}).

\bibitem{west2017scale}
\bibinfo{author}{West, G.~B.}
\newblock \emph{\bibinfo{title}{Scale: the universal laws of growth,
  innovation, sustainability, and the pace of life in organisms, cities,
  economies, and companies}} (\bibinfo{publisher}{Penguin},
  \bibinfo{year}{2017}).

\bibitem{bettencourt2007growth}
\bibinfo{author}{Bettencourt, L.~M.}, \bibinfo{author}{Lobo, J.},
  \bibinfo{author}{Helbing, D.}, \bibinfo{author}{K{\"u}hnert, C.} \&
  \bibinfo{author}{West, G.~B.}
\newblock \bibinfo{journal}{\bibinfo{title}{Growth, innovation, scaling, and
  the pace of life in cities}}.
\newblock {\emph{\JournalTitle{Proceedings of the national academy of
  sciences}}} \textbf{\bibinfo{volume}{104}}, \bibinfo{pages}{7301--7306}
  (\bibinfo{year}{2007}).

\bibitem{bonaventura21}
\bibinfo{author}{Bonaventura, M.}, \bibinfo{author}{Aiello, L.~M.},
  \bibinfo{author}{Quercia, D.} \& \bibinfo{author}{Latora, V.}
\newblock \bibinfo{journal}{\bibinfo{title}{{Predicting urban innovation from
  the US Workforce Mobility Network}}}.
\newblock {\emph{\JournalTitle{{Nature Humanities and Social Sciences
  Communications}}}} \textbf{\bibinfo{volume}{8}} (\bibinfo{year}{2021}).

\bibitem{richterich2014karma}
\bibinfo{author}{Richterich, A.}
\newblock \bibinfo{journal}{\bibinfo{title}{’karma, precious
  karma!’karmawhoring on reddit and the front page’s econometrisation}}.
\newblock {\emph{\JournalTitle{Journal of Peer Production}}}
  \textbf{\bibinfo{volume}{4}}, \bibinfo{pages}{1--12} (\bibinfo{year}{2014}).

\bibitem{liben2005geographic}
\bibinfo{author}{Liben-Nowell, D.}, \bibinfo{author}{Novak, J.},
  \bibinfo{author}{Kumar, R.}, \bibinfo{author}{Raghavan, P.} \&
  \bibinfo{author}{Tomkins, A.}
\newblock \bibinfo{journal}{\bibinfo{title}{Geographic routing in social
  networks}}.
\newblock {\emph{\JournalTitle{Proceedings of the National Academy of
  Sciences}}} \textbf{\bibinfo{volume}{102}}, \bibinfo{pages}{11623--11628}
  (\bibinfo{year}{2005}).

\bibitem{kuchler2021jue}
\bibinfo{author}{Kuchler, T.}, \bibinfo{author}{Russel, D.} \&
  \bibinfo{author}{Stroebel, J.}
\newblock \bibinfo{journal}{\bibinfo{title}{Jue insight: The geographic spread
  of covid-19 correlates with the structure of social networks as measured by
  facebook}}.
\newblock {\emph{\JournalTitle{Journal of Urban Economics}}}
  \bibinfo{pages}{103314} (\bibinfo{year}{2021}).

\bibitem{joglekar2020analysing}
\bibinfo{author}{Joglekar, S.}, \bibinfo{author}{Velupillai, S.},
  \bibinfo{author}{Dutta, R.} \& \bibinfo{author}{Sastry, N.}
\newblock \bibinfo{journal}{\bibinfo{title}{Analysing meso and macro
  conversation structures in an online suicide support forum}}.
\newblock {\emph{\JournalTitle{arXiv preprint arXiv:2007.10159}}}
  (\bibinfo{year}{2020}).

\bibitem{calvillo2020political}
\bibinfo{author}{Calvillo, D.~P.}, \bibinfo{author}{Ross, B.~J.},
  \bibinfo{author}{Garcia, R.~J.}, \bibinfo{author}{Smelter, T.~J.} \&
  \bibinfo{author}{Rutchick, A.~M.}
\newblock \bibinfo{journal}{\bibinfo{title}{Political ideology predicts
  perceptions of the threat of covid-19 (and susceptibility to fake news about
  it)}}.
\newblock {\emph{\JournalTitle{Social Psychological and Personality Science}}}
  \textbf{\bibinfo{volume}{11}}, \bibinfo{pages}{1119--1128}
  (\bibinfo{year}{2020}).

\bibitem{jhaver2019does}
\bibinfo{author}{Jhaver, S.}, \bibinfo{author}{Bruckman, A.} \&
  \bibinfo{author}{Gilbert, E.}
\newblock \bibinfo{journal}{\bibinfo{title}{Does transparency in moderation
  really matter? user behavior after content removal explanations on reddit}}.
\newblock {\emph{\JournalTitle{Proceedings of the ACM on Human-Computer
  Interaction}}} \textbf{\bibinfo{volume}{3}}, \bibinfo{pages}{1--27}
  (\bibinfo{year}{2019}).

\bibitem{mccrae1994stability}
\bibinfo{author}{McCrae, R.~R.} \& \bibinfo{author}{Costa~Jr, P.~T.}
\newblock \bibinfo{journal}{\bibinfo{title}{The stability of personality:
  Observations and evaluations}}.
\newblock {\emph{\JournalTitle{Current directions in psychological science}}}
  \textbf{\bibinfo{volume}{3}}, \bibinfo{pages}{173--175}
  (\bibinfo{year}{1994}).

\bibitem{schurer2015universities}
\bibinfo{author}{Schurer, S.}, \bibinfo{author}{de~New, S.} \&
  \bibinfo{author}{Leung, F.}
\newblock \bibinfo{title}{Do universities shape their students' personality?}
\newblock \bibinfo{type}{Tech. Rep.}, \bibinfo{institution}{Institute of Labor
  Economics (IZA)} (\bibinfo{year}{2015}).

\bibitem{bozarth2021market}
\bibinfo{author}{Bozarth, L.} \& \bibinfo{author}{Budak, C.}
\newblock \bibinfo{title}{Market forces: Quantifying the role of top credible
  ad servers in the fake news ecosystem}.
\newblock In \emph{\bibinfo{booktitle}{Proceedings of the International AAAI
  Conference on Web and Social Media}}, vol.~\bibinfo{volume}{15},
  \bibinfo{pages}{83--94} (\bibinfo{year}{2021}).

\bibitem{feingold2017fake}
\bibinfo{author}{Feingold, R.}
\newblock \bibinfo{title}{Fake news \& misinformation policy practicum}
  (\bibinfo{year}{2017}).

\bibitem{royal2022online}
\bibinfo{author}{{The Royal Society}}.
\newblock \bibinfo{title}{The online information environment: Understanding how
  the internet shapes people’s engagement with scientific information}.
\newblock \bibinfo{type}{Tech. Rep.}, \bibinfo{institution}{The Royal Society},
  \bibinfo{address}{London, UK} (\bibinfo{year}{2012}).

\bibitem{hindman2000rural}
\bibinfo{author}{Hindman, D.~B.}
\newblock \bibinfo{journal}{\bibinfo{title}{The rural-urban digital divide}}.
\newblock {\emph{\JournalTitle{Journalism \& Mass Communication Quarterly}}}
  \textbf{\bibinfo{volume}{77}}, \bibinfo{pages}{549--560}
  (\bibinfo{year}{2000}).

\end{thebibliography}












\end{document}